\documentclass[letterpaper, 10 pt, conference]{ieeeconf}  

\IEEEoverridecommandlockouts                              

\usepackage{amsfonts}
\usepackage{mathrsfs}
\usepackage{amssymb}
\usepackage{extarrows}
\usepackage{psfrag}
\usepackage{graphicx}
\usepackage{color}
\usepackage{cite}

\def\rep#1{(\ref{#1})}

\newcommand{\R}{\mathbb{R}}

\def\send#1#2{\stackrel{#1}{\hbox to #2{\rightarrowfill}}}
\def\-{\!\!\!\!\!-}

 \def\qed{ \rule{.08in}{.08in}}
\def\eq#1{\begin{equation}#1\end{equation}}

\def\scr#1{{\cal #1}}

\newtheorem{theorem}{Theorem}
\newtheorem{lemma}{Lemma}

\newtheorem{corollary}{Corollary}

\def\qed{ \rule{.1in}{.1in}}

\def\R{{\rm I\!R}} 

\newcounter{seqn}[equation]
\def\theseqn{\arabic{equation}\alph{seqn}}

\def\endseqn{\eqno \@seqnnum
$$\ignorespaces}
\def\@seqnnum{(\theseqn)}

\newskip\mcentering \mcentering=0pt plus 1000pt minus 1000pt

\def\meqalignno#1{
\halign to\displaywidth{
    \hbox to 0pt{\kern\displaywidth\llap{$##$}\hss}\tabskip=\mcentering
    &\hfil$\displaystyle{##}$\tabskip=\mcentering
   &&$\displaystyle{{}##}$\hfil\tabskip=\mcentering
    \crcr
    #1\crcr}}

\def\rep#1{(\ref{#1})}

\def\eq#1{\begin{equation}#1\end{equation}}

\def\dspace{\multiply\normalbaselineskip 150
          \divide\normalbaselineskip 100 \normalbaselines
          \csname @@normalbaselineskip\endcsname\normalbaselineskip}
\def\sspace{\multiply\normalbaselineskip 200
         \divide\normalbaselineskip 300 \normalbaselines
         \csname @@normalbaselineskip\endcsname\normalbaselineskip}
\def\sdspace{\multiply\normalbaselineskip 160
         \divide\normalbaselineskip 150 \normalbaselines
         \csname @@normalbaselineskip\endcsname\normalbaselineskip}

\def\@{\tilde}

\def\3dot#1{\buildrel\textstyle...\over#1}

\title{\LARGE \bf
A Distributed Observer for a Continuous-Time Linear System with time-varying network
}

\author{Lili Wang, Ji Liu, and  A. Stephen Morse
\thanks{L. Wang and A.S. Morse  are with the Department
of Electrical Engineering, Yale University
({\tt\small \{lili.wang,  as.morse\}@yale.edu}).
J.~Liu is with the Department of Electrical and Computer Engineering, Stony Brook University
({\tt\small ji.liu@stonybrook.edu}).
}%
}

\begin{document}
\maketitle
\thispagestyle{empty}

\begin{abstract}

A simply structured distributed  observer is described for estimating the state of a continuous-time, jointly observable, input-free, linear system whose sensed outputs are distributed across a time-varying network.
It is explained how to design a gain $g$ in the observer so that their state estimation errors all converge exponentially fast to zero at a fixed, but arbitrarily chosen rate provided the network's graph is  strongly connected for all time.
A linear inequality for $g$ is provided when the network's graph is switching according to a switching signal with a dwell time or an average dwell time, respectively.
It has also been shown the existence of $g$  when the stochastic matrix of the network's graph is chosen to be doubly stochastic under arbitrarily switching signals.
This is accomplished by  exploiting several well-known properties of invariant subspaces and properties of perturbed systems.
\end{abstract}

\section{Introduction}
Distributed state estimation problem has gotten more and more attention in recent years\cite{reza-cdc2005,reza-cdc2007, reza-tac2012,  martins-cdc2012, martins-tac2017,trentlemen-scl2018, trentlemen-tac2019,Mitra-tac2018,Mitra-acc2019,CDC17,TAC18,kim-cdc2016,Kim-tac2019} due to the increasing interest in sensor network and multi-agent systems.
The problem is to enable each agent to reconstruct the system state by using its own measurements and communicating with the nearby neighbors in a network. 
More specifically, for the continuous time case it is to estimate the state of an $m> 0$ channel, $n$-dimensional continuous-time linear system of the form $\dot x= Ax, \; y_i=C_i x,\; i\in \{1,2,\ldots,m\}$ under the necessary assumption that the system is ``jointly observable''. 
This problem has been studied in different forms. 
This problem is originally studied in \cite{reza-cdc2005,reza-cdc2007, reza-tac2012} through a consensus-based Kalman filter assuming that data fusion can be achieved in finite time.
In \cite{martins-cdc2012, martins-tac2017, TAC18}, this problem is solved by recasting this as a decentralized control problem. 
The method allows to freely assign the spectrum of the estimators under the condition that the network is strongly connected and fixed.
The recent work in \cite{Mitra-tac2018,Mitra-acc2019} studies this problem based on the structure of the network. 
By choosing or constructing a tree in the network, it is able to broadcast the information and  do estimation. 

In this paper, a simple distributed observer is designed by exploiting several well-known properties of invariant subspaces, i.e., the properties of the unobservable subspaces of each agent.
The idea stems from research originally  reported in \cite{kim-cdc2016,Kim-tac2019} and  subsequently extended in \cite{trentlemen-tac2019}. 
This simplified observer is described and its behaviors is analyzed in \cite{ACC19}, and a discrete-time version is studied in \cite{CDC19}.
As they stand, those estimators in \cite{kim-cdc2016,Kim-tac2019,trentlemen-tac2019,trentlemen-scl2018} can only deal with the case when the neighbor graph is fixed.
The aims of this paper is to extend the results in \cite{ACC19} to the case when the network's graph is time-varying.

\subsection{Invariant Subspaces}\label{sec:invariant}
Throughout this paper certain basic and well-known algebraic properties of invariant subspaces will be exploited.
To understand what they are, let $A$ be any square matrix, and suppose $\mathcal{V}$ is an $A$-invariant subspace. Let $Q$  be any full row rank matrix whose kernel is $\mathcal{V}$ and  suppose that  $V$ is any ``basis matrix''
 for $\mathcal{V}$; i.e., a matrix whose columns form a basis for $\mathcal{V}$.
Then the linear equations
\[QA =\bar{A}_{V}Q\;\;\;\;\; \text{and}\;\;\;\;\; AV=VA_{V}\]
  have unique solutions $\bar{A}_V$ and $A_V$ respectively. 
Let $V^{-1}$ be any left inverse of $V$ and let
    $Q^{-1}$ be that right inverse of $Q$ for which $V^{-1}Q^{-1} =0 $.
    Then
\[A = H^{-1}\begin{bmatrix}\bar{A}_V & 0\\ \widehat{A}_V & A_{V}\end{bmatrix}H,\;\; \text{
 where }\;\;
 H = \begin{bmatrix}Q \\ V^{-1}\end{bmatrix}\]
and
 $\widehat{A}_V = V^{-1}AQ^{-1}$.
Use will be made of these simple algebraic facts in the sequel.

\section{Problem} We are interested in a  network of $m>1$ agents labeled $1,2,\ldots,m$ which are able to receive information from their neighbors where by a {\em neighbor} of agent $i$ is meant
any agent in agent $i$'s reception range. In this paper, the network we consider is time-varying and changes according to a switching signal.
Let $\scr{P}$ denote a suitably defined set, indexing the set of all possible network.
Let $\tau_D>0$ and let $\sigma:[0,\infty)\rightarrow\scr{P}$ be a piecewise-constant switching signal
whose switching times $t_1,t_2,\ldots$ satisfy $t_{i+1}-t_i\geq\tau_D$, $i\geq 0$.


We write $\mathcal{N}_i(\sigma(t))$ for the set of labels of agent $i$'s neighbors at time $t$ and take agent $i$ to be a neighbor of itself. Relations between neighbors are characterized by a directed graph
  $\mathbb{N}(\sigma(t))$ with $m$ vertices and a set of arcs defined so that there is an arc from vertex $j$ to vertex $i$ whenever
  agent $j$ is a neighbor of agent $i$. 
    Each agent $i$ can sense a
  signal $y_i\in\R^{s_i}$ where
\eq{\dot{x}=Ax,\;\;\;y_i = C_ix,\;\;\;i\in\mathbf{m} = \{1,2,\ldots,m\}\label{eq:syss},}
and $x\in\R^n$. 
We assume throughout that $\mathbb{N}(\sigma(t))$ is strongly connected and
that the system defined by \rep{eq:syss} is {\em jointly observable}; i.e., with
$C = \begin{bmatrix}C_1' &C_2' & \cdots & C_m'\end{bmatrix}'$, the matrix pair $(C,A)$ is observable. Joint observability is
equivalent to the requirement that
$$\bigcap_{i\in\mathbf{m}}\scr{V}_i = 0$$
where $\scr{V}_i$ is the {\em unobservable space} of $(C_i,A)$. 
As is well known, $\scr{V}_i$ is
 the largest $A$-invariant subspace contained in the
kernel of $C_i$.

Each agent $i$  is to  estimate $x$ using an $n$-dimensional linear system with state $x_i\in\R^n$ and we assume that the information
agent $i$ can receive from neighbor $j$ at time $t$ is $x_j(t)$.
The problem of interest is to construct a suitably defined family of linear  estimators in such a way so that under switching neighbor graphs no matter what the  estimators' initial states are, for each
 $i\in\mathbf{m}$,
 $x_i(t)$ is an asymptotically correct estimate of $x(t)$ in the sense that the estimation error
$x_i(t)-x(t)$ converges to zero  as fast as $e^{-\lambda t}$ does, where $\lambda$ is an
arbitrarily chosen but fixed
 positive number.

\section{The Observer}

The observer to be considered consists of $m$ private estimators  of the form
\begin{eqnarray}\dot{x}_i &= & (A+K_iC_i)x_i -K_iy_i
\nonumber\\&\ & -gP_i\bigg (x_i-\frac{1}{m_i(\sigma(t))}\sum_{j\in\mathcal{N}_i(\sigma(t)
)} x_j \bigg),\;\;i\in\mathbf{m}\label{eq:observer}\end{eqnarray}
where $m_i(\sigma(t))$ is the number of labels in $\mathcal{N}_i(\sigma(t))$, $g$ is a suitably defined positive gain, each $K_i$ is a suitably defined matrix, and for each $i\in\mathbf{m}$,  $P_i$ is the orthogonal projection on the unobservable space of $(C_i,A)$.

To begin with, each matrix  $K_i$ is defined as follows.
For each fixed $i\in\mathbf{m}$,  write $Q_i$ for any full  rank matrix whose kernel is the unobservable space of $(C_i,A)$ and let $\bar{C}_i$ and $\bar{A}_i$ be the unique solutions to $\bar{C}_iQ_i = C_i$ and $Q_iA=\bar{A}_iQ_i$  respectively.  Then  the matrix pair $(\bar{C}_i,\bar{A}_i)$ is observable.  
Thus by using a standard spectrum assignment algorithm, a matrix $\bar{K}_i$ can be chosen to ensure that  the convergence of $e^{(\bar{A}_i + \bar{K}_i\bar{C}_i)t}$ to zero is as fast as the convergence of $e^{-\hat \lambda t}$ to zero is. Here $\hat \lambda$ is a positive number which is greater than $\lambda$.
Having chosen such $\bar{K}_i$, $K_i$
is then chosen to be $K_i = Q_i^{-1}\bar{K}_i$ where $Q_i^{-1}$ is a right inverse for $Q_i$.
The definition implies that $Q_i(A+K_iC_i) = (\bar{A}_i+\bar{K}_i\bar{C}_i)Q_i$ and that
$(A+K_iC_i)\scr{V}_i\subset \scr{V}_i$. The latter, in turn, implies that
there is a unique matrix  $A_i$ which satisfies $(A+K_iC_i)V_i = V_iA_i$ where $V_i$ is a basis
matrix\footnote{For simplicity, we assume that the columns of $V_i$ constitute an orthonormal basis for $\scr{V}_i$ in which case  $P_i = V_iV_i'$.} for $\scr{V}_i$.
        To understand what needs to
         be considered in choosing
        $g$ it is necessary to delve more deeply into the structure of the overall observer. This will be done next.


\section{Analysis}

For each $i\in\mathbf{m}$, write $e_i$  for the {\em state estimation error} $e_i = x_i-x$. In view of \rep{eq:syss} and \rep{eq:observer},
\begin{eqnarray}\dot{e}_i & = & (A+K_iC_i)e_i \nonumber\\ &\ & - gP_i\left (e_i-\frac{1}{m_i(\sigma(t))}\sum_{j\in\scr{N}_i(\sigma(t))}e_i\right  )\label{eq:err}\end{eqnarray}
It is possible to combine these $m$ error equations into a single equation
 with state  $e = $ column $\{e_1,e_2,\ldots, e_m\}$. For this let
$\bar{A} = $ block diagonal $\{A+K_1C_1 ,A+K_2C_2,\ldots,A+K_mC_m\}$,
 $P = $ block diagonal $\{P_1 ,P_2,\ldots, P_m\}$ and write $S(\sigma(t))$ for the stochastic matrix
$S(\sigma(t)) = D_{\mathbb{N}(\sigma(t))}^{-1}A'_{\mathbb{N}(\sigma(t))}$ where $A_{\mathbb{N}(\sigma(t))}$ is the adjacency matrix of $\mathbb{N}(\sigma(t))$ and  $ D_{\mathbb{N}(\sigma(t))}$
 is the diagonal matrix whose $i$th diagonal entry is the in-degree
  of $\mathbb{N}(\sigma(t))$'s $i$th vertex.
The error model  is then
\eq{\dot{e} = (\bar{A} -gP((I_m-S(\sigma(t)))\otimes I_n))e\label{eq:em}}
where $\otimes$ denotes the Kronecker product.

 As a first step towards this end, note that for any value of $g$,  the direct sum
$\scr{V} = \scr{V}_1\oplus\scr{V}_2\oplus \cdots \oplus \scr{V}_m$
is $\bar{A} -gP((I_m-S(\sigma(t)))\otimes I_n)$ invariant. This is because $(A+K_iC_i)\scr{V}_i\subset \scr{V}_i,\;i\in\mathbf{m}$ and because
$\scr{V} = $ column span of $P$. Let $Q = $ block diagonal $ \{Q_1, Q_2,\ldots ,Q_m\}$
and $V = $ block diagonal $ \{V_1, V_2,\ldots ,V_m\}$
in which case
$Q$ is a full rank matrix whose kernel is $\scr{V}$  and $V$ is a basis matrix for $\scr{V}$ whose
 columns form an orthonormal set. It follows that $P = VV'$, that
$Q(\bar{A} -gP((I_m-S(\sigma(t)))\otimes I_n)) = ($block diagonal $\{\bar{A}_1+\bar{K}_1\bar{C}_1 ,\bar{A}_2+\bar{K}_2\bar{C}_2,\ldots,
\bar{A}_m+\bar{K}_m\bar{C}_m\})Q$ and that $(\bar{A} -gP((I_m-S(\sigma(t)))\otimes I_n))V = V(\tilde{A} -gV'((I_m-S(\sigma(t)))\otimes I_n)V)$
where $\tilde{A}$ is the unique solution to $\bar{A}V = V\tilde{A}$.
Let $V^{-1}$ be any left inverse of $V$ and let $Q^{-1}$ be that right inverse of $Q$ for which $V^{-1}Q^{-1} =0 $ and $V'=V^{-1}$.
Then

\begin{eqnarray*}
&\ &\bar{A} -gP((I_m-S(\sigma(t)))\otimes I_n) \\& =& H^{-1}\begin{bmatrix}\bar A_V& 0\\ \widehat{A}_V (\sigma(t)) & A_{V}(\sigma(t))\end{bmatrix}H,
\end{eqnarray*}
 where
\[ H = \begin{bmatrix}Q \\ V'\end{bmatrix},\]\[
 \bar A_V=\begin{bmatrix}\bar{A}_1+\bar{K}_1\bar{C}_1 & 0 & \ldots & 0\\ 0 & \bar{A}_2+\bar{K}_2\bar{C}_2 & \ldots & 0\\ \vdots & \vdots &\ddots & \vdots \\0 & 0 & \ldots & \bar{A}_m+\bar{K}_m\bar{C}_m\end{bmatrix},\]
 $A_V(\sigma(t))=\tilde{A} -gV'((I_m-S(\sigma(t)))\otimes I_n)V$, and 
 $\widehat{A}_V (\sigma(t))= V'\bar AQ^{-1}-gV'((I_m-S(\sigma(t)))\otimes I_n)Q^{-1}$.
 
 In order to show exponential convergence of $x_i(t)-x(t)$, i.e., \eqref{eq:em}, it is equivalent to look at the stability of system
 \begin{equation}\label{eq:new}
     \begin{bmatrix}\dot{z}_1\\\dot{z}_2\end{bmatrix}=\begin{bmatrix}\bar A_V& 0\\ \widehat{A}_V (\sigma(t)) & A_{V}(\sigma(t))\end{bmatrix}  \begin{bmatrix}{z}_1\\{z}_2\end{bmatrix}
 \end{equation}
 where $e=H\begin{bmatrix}z_1'&z_2'\end{bmatrix}'$.
System~\eqref{eq:new} can be written as  
 \begin{equation}\label{eq:z1}
\dot{z}_1=\bar A_V z_1\end{equation}
and 
\begin{equation}\label{eq:z2}
    \dot{z}_2= A_V (\sigma(t))z_2+\hat{A}_V(\sigma(t))z_1
\end{equation}

\subsection{Stability of a switching system with dwell time $\tau_D$}

  Notice that in practical applications, the neighbor graph usually does not switch arbitrarily fast. 
In other words, the change of neighbor graphs must satisfy a dwell time constraint or an average dwell time constraint.
Given a positive constant $\tau_D$, let $\mathcal S(\tau_D)$ denote the set of
all switching signals with interval between consecutive discontinuities no smaller than  $\tau_D$. The constant  $\tau_D$ is called {\it the (fixed) dwell-time. } The following result is developed in the paper.

\begin{theorem} For any switching signal $\sigma(t)\in \mathcal S(\tau_D): \mathbb{R}\rightarrow \mathcal{P}$ with any dwell time $\tau_D>0$, and any given positive number $\lambda$, if the neighbor graph $\mathbb{N}(\sigma(t))$ is strongly connected and the system defined by \eqref{eq:syss} is jointly observable, there are matrices $K_i,\;i\in\mathbf{m}$ such that for $g$
sufficiently large, each state estimation error $x_i(t)-x(t) $ of the  distributed observer defined by \eqref{eq:observer},
 converges to zero
as  $t\rightarrow \infty$ as fast as $e^{-\lambda t}$ converges to zero. \label{thm:1}\end{theorem}

By \cite{ACC19} if $\sigma(t)$ is a fixed constant, the estimation error $e(t)$ of \eqref{eq:em} can converge to zero as fast as $e^{-\lambda t}$  by choosing large enough $g$.  
However, the stability of all subsystems \eqref{eq:em} for each fixed value of $\sigma(t)$ does not ensure the stability of the switching system. 
Two lemmas about switching system are needed in order to prove theorem~\ref{thm:1}.

\begin{lemma}\label{lemma:g}
Given a set of matrices  $\{M(1),\; M(2),\;\ldots, \; M(p)\}$ where each $M(i)\in \mathbb{R}^{n\times n}$ $i\in \mathcal{P}$ is exponentially stable. 
For any $\tau_D>0$, and $\sigma(t)\in \mathcal S(\tau_D) : \mathbb{R}\rightarrow \mathcal{P} $, there exists a positive number $g$
so that $x$ converges to zero as fast as a preassigned convergence rate $\lambda$ under the switching system
\[\dot{x}=gM(\sigma(t))x\]
\end{lemma}

\noindent \textbf{Proof} 
Since $M(i)$ $i\in \mathcal{P}$ is exponentially  stable, there is are positive constants $c_i$ and $\lambda_i$ such that
\[\|e^{M(i)t}\|\leq c_i e^{-\lambda_i t}\]
where $\| \cdot \|$ is any norm on $\mathbb{R}^{n\times n}$ for which
the submultiplicative property holds.
Each $c_i$ is chosen to be larger than $1$.
Thus,
\[\|e^{gM(i)t}\|=\|e^{M(i)gt}\|\leq c_i e^{-\lambda_i gt}\]
By \cite[Lemma 2]{Morse-tac1996}, for $g\lambda_i\geq \lambda$ if \[\tau_D\geq \frac{\ln c_i}{g\lambda_i-\lambda} \;\;\;\;  \forall i\in \mathcal{P}\]
$x(t)\leq c e^{-\lambda t}x(0)$  where $c=\max_{i\in \mathcal{P}} c_i$.
Therefore, by choosing $g$ so that 
$g\lambda_i\geq \lambda$, and 
$g\geq \frac{\ln c_i+\lambda \tau _D}{\lambda_i\tau_D}$ $\forall i\in \mathcal{P}$, for any given $\sigma(t)$ with dwell time $\tau_D$ there exists $g$ so that $x$ converges to zero with convergence rate $\lambda$.
$\qed$

\begin{lemma}\label{lemma:gA+B}
Given a set of matrices  $\{M(1),\; M(2),\;\ldots, \; M(p)\}$ where each $M(i)\in \mathbb{R}^{n\times n}$ $i\in \mathcal{P}$ is exponentially stable, and a bounded matrix $N\in \mathbb{R}^{n\times n}$. 
For any $\tau_D>0$, and $\sigma(t) \in \mathcal S(\tau_D) : \mathbb{R}\rightarrow \mathcal{P}$, there exists a positive number $g$
so that $x$ converges to zero as fast as a preassigned convergence rate $\lambda$ under the switching system
\[\dot{x}=\left(N+gM(\sigma(t))\right)x\]
\end{lemma}

\noindent \textbf{Proof} 
Since $M(i)$ $i\in \mathcal{P}$ is exponentially  stable, there is are positive constants $c_i$ and $\lambda_i$ such that
\[\|e^{M(i)t}\|\leq c_i e^{-\lambda_i t}\]
where $\| \cdot \|$ is any norm or induced norm on $\mathbb{R}^{n\times n}$ for which
the submultiplicative property holds.
Each $c_i$ is chosen to be larger than $1$.
Let $\Phi_{gM}(t,\tau)$ be the transition matrix of $gM(\sigma(t))$.
By Lemma~\ref{lemma:g}, for any given  positive number $\bar \lambda$, and $\sigma(t)$ with dwell time $\tau_D$ by choosing $g$ so that $g\lambda_i\geq \bar \lambda$, and 
$g\geq \frac{\ln c_i+\bar \lambda \tau _D}{\lambda_i\tau_D}$ $\forall i\in \mathcal{P}$, there exists $g$ so that $\Phi_{gM}(t,\tau)\leq c e^{-\bar \lambda(t-\tau)} $ where $c=\max_{i\in \mathcal{P}} c_i$.

Since $N$ is bounded, there exists $b$ so that $\|N\|\leq b$.

Now look at system $\dot{x}=gM(\sigma(t))x+Nx$. 
By viewing $Nx$ as a forcing function in the preceding, one may write the variation of constants formula
\[x(t)=\Phi_{gM}(t,0) x(0)+\int_0^t\Phi_{gM}(t,\mu)Nx(\mu)d\mu
\]
Therefore
\begin{eqnarray*}
\|x(t)\| & \leq & \|\Phi_{gM}(t,0)x(0)\|+\int_0^t\|\Phi_{gM}(t,\mu)Nx(\mu)\|d\mu\\
&\leq & c e^{-\bar \lambda t} \|x(0)\| +\int_0^tc e^{-\bar \lambda (t-\mu)} \|N\|\|x(\mu)\|d\mu
\end{eqnarray*}
That is 
\[e^{\bar \lambda t} \|x(t)\|  \leq   c \|x(0)\| +\int_0^t bc e^{\bar \lambda \mu} \|x(\mu)\|d\mu
\]
By the Bellman-Gronwall Lemma,
\[e^{\bar \lambda t} \|x(t)\|  \leq  c \|x(0)\| e^{\int_0^t bc d\mu }=c\|x(0)\| e^{bct }
\]
Therefore 
\[\|x(t)\|\leq c \|x(0)\| e^{(bc-\bar \lambda)t }
\]
If $\bar \lambda$ is chosen $\bar \lambda \geq \lambda+bc$, and  $g$ is chosen so that $g\lambda_i\geq \bar \lambda$, and 
$g\geq \frac{\ln c_i+\bar \lambda \tau _D}{\lambda_i\tau_D}$ $\forall i\in \mathcal{P}$,
\[\|x(t)\|\leq c \|x(0)\| e^{-\lambda t }.
\]
$\qed$

Proof of Theorem~\ref{thm:1} is provided in the following.

\noindent \textbf{Proof of Theorem \ref{thm:1}}
In order to show exponential convergence of $x_i(t)-x(t)$, i.e., \eqref{eq:em}, it is equivalent to look at the stability of system~\eqref{eq:new}, i.e., \eqref{eq:z1} and \eqref{eq:z2}.
First, since the spectrum of  $\bar{A}_i+\bar{K}_i\bar{C}_i,\;i\in\mathbf{m}$, is assignable with  $\bar{K}_i$, there are $
\bar K_i$s so that $\|z_1(t)\|\leq   e^{-\hat \lambda t} \|z_1(0)\|$ where $\hat \lambda> \lambda$.
It is left to show that for $g$ sufficiently large,
 $z_2$ of \eqref{eq:z2} converges to zero with a prescribed convergence rate as large as $ \lambda$.
 
Consider $\hat A_V(\sigma(t))z_1$ of \eqref{eq:z2} as a forcing function, and let $\Phi_{V}(t,\tau)$ be the transition matrix of $A_V(\sigma(t))$ for any $t\geq \tau \geq 0$. Thus,
\[
z_2(t)=\Phi_V(t,0)z_2(0)+\int_0^t\Phi_V(t,\mu)\hat A_V(\sigma(\mu))z_1(\mu)d\mu
\]
Recall that $A_V(\sigma(t))=\tilde{A} -gV'((I_m-S(\sigma(t)))\otimes I_n)V$. 
By \cite[Proposition 1]{ACC19}, for any $i\in \mathcal{P}$,
$-V'((I_m-S(i))\otimes I_n)V$ is exponentially stable.
Let $c_i$ and $\lambda_i$ be two positive constants such that
\[\|e^{-V'((I_m-S(i))\otimes I_n)Vt}\|\leq c_i e^{-\lambda_i t}\]
Each $c_i$ is chosen to be larger than $1$.
Let $c=\max_{i\in \mathcal{P}} c_i$.
Since $\tilde{A}$ is fixed, let $
\|\tilde{A}\|\leq b$. Moreover
according to Lemma~\ref{lemma:gA+B}, 
if for $\bar \lambda \geq \lambda+ b c$, $g$ is chosen so that $g\lambda_i\geq \bar \lambda$, and 
$g\geq \frac{\ln c_i+\bar \lambda \tau _D}{\lambda_i\tau_D}$ $\forall i\in \mathcal{P}$,

\[\|\Phi_V(t,\tau)\|\leq c e^{-\lambda(t-\tau)},\;\;\; \forall t\geq \tau\geq 0\]
Once $g$ is fixed, there exists a positive number $\hat c$ so that 
$\|\hat A_V(\sigma(\mu))\|
\leq \hat c$.
Therefore, 
 \begin{eqnarray*}
 \|z_2(t)\|& \leq & \|\Phi_V(t,0)\| \|z_2(0)\|\\ &\ & +\int_0^t\|\Phi_V(t,\mu)\| \|\hat A_V(\sigma(\mu))\|\|z_1(\mu)\|d\mu
 \\ &\leq & 
 c e^{-\lambda t}\|z_2(0)\| +\int_0^t c e^{-\lambda (t-\mu)} \hat c e^{-\hat\lambda \mu}\|z_1(0)\|d\mu \\ &=& c e^{-\lambda t}\|z_2(0)\| +c \hat c e^{-\lambda t}\|z_1(0)\|\int_0^t  e^{(\lambda-\hat \lambda) \mu} d\mu
 \\ &=& 
 c e^{-\lambda t}\|z_2(0)\| -c \hat c \frac{1}{\hat \lambda- \lambda}\|z_1(0)\|e^{-\hat \lambda t}
 \\ &\ & +c \hat c \frac{1}{\hat \lambda- \lambda} \|z_1(0)\|e^{-\lambda t} 
 \\ & \leq &  c e^{-\lambda t}\|z_2(0)\| 
 +c \hat c \frac{1}{\hat \lambda- \lambda} \|z_1(0)\|e^{-\lambda t} 
 \end{eqnarray*}
 Thus 
  \[
  \begin{bmatrix}
  \|z_1(t)\|\\ \|z_2(t)\|
  \end{bmatrix}\leq e^{-\lambda t}\begin{bmatrix} 1 & 0 \\ c\hat c \frac{1}{\hat \lambda-\lambda} & c\end{bmatrix}\begin{bmatrix}\|z_1(0)\|\\\|z_2(0)\|
  \end{bmatrix}
  \]
$\qed$

According to the proof of Theorem~\ref{thm:1}, the bound of $g$ can be derived.
Recall that  $c_i$ and $\lambda_i$ are two positive constants such that
\[\|e^{-V'((I_m-S(i))\otimes I_n)Vt}\|\leq c_i e^{-\lambda_i t}\]
and $c=\max_{i\in \mathcal{P}} c_i$.
$b$ is the constant so that$
\|\tilde{A}\|\leq b$.
Let $\lambda^*=\min_{i\in \mathcal{P}} \lambda_i$.
\begin{equation}\label{eq:g}
    g\geq \max \{\frac{\lambda +bc}{\lambda^*},\frac{\lambda +bc}{\lambda^*}+\frac{\ln c}{\lambda^* \tau_D}\}
\end{equation}

\subsection{Stability of a  switching system with average dwell time $\tau_D$}

However, in certain situations, the switching signals may occasionally have consecutive discontinuities separated by less than $\tau_D$, but for which the average interval between consecutive discontinuities is no less than $\tau_D$. This leads to the concept of average dwell time. 
For each switching signal $\sigma(t)$ and each $t \geq \tau \geq 0$, let $N_{\sigma}(t,\tau)$ denote the number of discontinuities of $\sigma(t)$ in the open interval $(\tau, t)$. For given $N_0, \tau_D > 0$, we
denote by $\mathcal{S}_{ave}(\tau_D,N_0)$ the set of all switching signals
for which
\[
 \mathcal{S}_{ave}(\tau_D,N_0)=\{\sigma(t): N_{\sigma}(t_0,t)\leq N_0+\frac{t-t_0}{\tau_D}  \}.
\]
The constant $\tau_D$ is called {\it the average dwell-time} and
$N_0$ {\it the chatter bound.} Correspondingly the following result is developed.

\begin{corollary} For any switching signal  $\sigma(t)\in \mathcal S_{ave}(\tau_D, N_0): \mathbb{R}\rightarrow \mathcal{P}$ with any average dwell time $\tau_D>0$ and arbitrary chatter bound $N_0$, and any given positive number $\lambda$, if the neighbor graph $\mathbb{N}(\sigma(t))$ is strongly connected and the system defined by \eqref{eq:syss} is jointly observable, there are matrices $K_i,\;i\in\mathbf{m}$ such that for $g$
sufficiently large, each state estimation error $x_i(t)-x(t) $ of the  distributed observer defined by \eqref{eq:observer}, converges to zero as  $t\rightarrow \infty$ as fast as $e^{-\lambda t}$ converges to zero. \label{corollary}\end{corollary}

In order to prove Corollary \ref{corollary}, the following two lemmas which are the switching signal with average dwell time version of Lemma \ref{lemma:g} and Lemma \ref{lemma:gA+B}, respectively.

\begin{lemma}\label{lemma:g2}
Given a set of matrices  $\{M(1),\; M(2),\;\ldots, \; M(p)\}$ where each $M(i)\in \mathbb{R}^{n\times n}$ $i\in \mathcal{P}$ is exponentially stable.
For any switching signal  $\sigma(t)\in \mathcal S_{ave}(\tau_D, N_0)$ with any average dwell time $\tau_D>0$ and arbitrary chatter bound $N_0$, there exists a positive number $g$
so that $x$ converges to zero as fast as a preassigned convergence rate $\lambda$ under the switching system
\[\dot{x}=gM(\sigma(t))x\]
\end{lemma}

\noindent \textbf{Proof} 
Since $M(i)$ $i\in \mathcal{P}$ is exponentially  stable, there is are positive constants $c_i$ and $\lambda_i$ such that
\[\|e^{M(i)t}\|\leq c_i e^{-\lambda_i t}\]
where $\| \cdot \|$ is any norm on $\mathbb{R}^{n\times n}$ for which
the submultiplicative property holds.
Thus,
\[\|e^{gM(i)t}\|=\|e^{M(i)gt}\|\leq c_i e^{-\lambda_i gt}\]
Each $c_i$ is chosen to be larger than $1$.
Let  $c=\max_{i\in \mathcal{P}} c_i$, 
and $\lambda^*=\min_{i\in \mathcal{P}} \lambda_i$.
Thus \[\|e^{gM(i)t}\|\leq c e^{-\lambda^* gt}\]
Let $\Phi_M(t,\tau)$ be the state transition matrix of system $\dot{x}=gM(\sigma(t))x$, 
\[\|\Phi_M(t,0)\|\leq  c^{N_{\sigma}(0,t)}e^{-g\lambda^*t}
\]
where $N_{\sigma}(0,t)\leq N_0+\frac{t}{\tau_D}  $  is the number of switching between $(0,t)$.
Choose $g$, so that $g\geq \frac{\ln c+\lambda \tau _D}{\lambda^*\tau_D}$.
Then we have 
\[\tau_D\geq \frac{\ln c}{g\lambda^*-\lambda} \;\;\;\;  \forall i\in \mathcal{P}\]
Therefore
\[
\|\Phi_M(t,0)\|\leq  c^{N_0}c^{\frac{t}{\tau_D}}e^{-g\lambda^*t}\leq c^{N_0}e^{-\lambda t} 
\]

Therefore, by choosing $g$ so that 
$g\lambda_i\geq \lambda$, and 
$g\geq \frac{\ln c+\lambda \tau _D}{\lambda^*\tau_D}$, for any given $\sigma(t)$ with average dwell time $\tau_D$ there exists $g$ so that $x$ converges to zero with convergence rate $\lambda$.
$\qed$

\begin{lemma}\label{lemma:gA+B2}
Given a set of matrices  $\{M(1),\; M(2),\;\ldots, \; M(p)\}$ where each $M(i)\in \mathbb{R}^{n\times n}$ $i\in \mathcal{P}$ is exponentially stable, and a bounded matrix $N\in \mathbb{R}^{n\times n}$. 
For any switching signal  $\sigma(t)\in \mathcal S_{ave}(\tau_D, N_0)$ with any average dwell time $\tau_D>0$ and arbitrary chatter bound $N_0$, there exists a positive number $g$
so that $x$ converges to zero as fast as a preassigned convergence rate $\lambda$ under the switching system
\[\dot{x}=\left(N+gM(\sigma(t))\right)x\]
\end{lemma}
By using the result of Lemma~\ref{lemma:g2},  the proof of Lemma~\ref{lemma:gA+B2} is similar to the proof of Lemma~\ref{lemma:gA+B} which is omitted here.

Then based on Lemma~\ref{lemma:gA+B2},
the proof of Corollary \ref{corollary} is almost the same as the proof of Theorem~\ref{thm:1} which is omitted here.
The linear inequality which $g$ has to satisfy to ensure the convergence rate of \eqref{eq:em} is still \eqref{eq:g} where $\tau_D$ will be the average dwell time instead of dwell time.

\subsection{Stability of a switching system with arbitrary switching}
In this section, a special case is studied. It can be shown that when   the stochastic matrix $S(\sigma(t))$ of each neighbor graph $\mathbb{N}(\sigma(t))$ is doubly stochastic, state estimation can be achieved under arbitrary switching.

\begin{theorem} For any switching signal $\sigma(t): \mathbb{R}\rightarrow \mathcal{P}$, and any given positive number $\lambda$, if the neighbor graph $\mathbb{N}(\sigma(t))$ is strongly connected,  the stochastic matrix $S(\sigma(t))$ of graph $\mathbb{N}(\sigma(t))$ is doubly stochastic, and the system defined by \eqref{eq:syss} is jointly observable, there are matrices $K_i,\;i\in\mathbf{m}$ such that for $g$
sufficiently large, each state estimation error $x_i(t)-x(t) $ of the  distributed observer defined by \eqref{eq:observer},
 converges to zero
as  $t\rightarrow \infty$ as fast as $e^{-\lambda t}$ converges to zero. \label{thm:2}\end{theorem}

\noindent {\bf Proof:} 
Recall $\Phi_{V}(t,\tau)$ is the transition matrix of $A_V(\sigma(t))$ for any $t\geq \tau \geq 0$.
If we can show that there exist a constant $c$ so that
\[\|\Phi_V(t,\tau)\|\leq c e^{-\lambda(t-\tau)},\;\;\; \forall t\geq \tau\geq 0\]
the remaining proof is exactly the same as the proof of Theorem~\ref{thm:1} which is omitted here. 

It is left to show that $\|\Phi_V(t,\tau)\|\leq c e^{-\lambda(t-\tau)},\;\;\; \forall t\geq \tau\geq 0$ by choosing $g$ sufficiently large.
We exploit matrix $A_V(\sigma(t))$.
Recall that $A_V(\sigma(t))=\tilde{A}-gV'((I_m-S(\sigma(t)))\otimes I_n)V$. 
In particular, 
\begin{eqnarray*}
&\ & (\lambda I + A_V(\sigma(t)))+(\lambda I + A_V(\sigma(t)))'\\ &=&
(\lambda I+\Tilde{A})+(\lambda I+\Tilde{A})'\\ &\ &-gV'((2I_m-S(\sigma(t))-S'(\sigma(t)))\otimes I_n)V
\end{eqnarray*}
Since each $S(\sigma(t))$ is doubly stochastic, $2I_m-S(\sigma(t))-S'(\sigma(t))$ has row sum $0$,  all its off-diagonal entries are non-positive, and all its diagonal entries are positive. 
That is this matrix can be seen as a generalized Laplacian matrix of a strongly connected graph.
By \cite[Proposition 1]{ACC19}, for any $t$,
$-V'((2I_m-S(\sigma(t))-S'(\sigma(t)))\otimes I_n)V$ is negative definite.
Thus by picking $g$ sufficiently large,  $ (\lambda I + A_V(\sigma(t)))+(\lambda I + A_V(\sigma(t)))'$ will be negative definite for any time $t$.

Consider system 
\[\dot {\bar z}=A_V(\sigma(t)) \bar z\]
Let $V=\bar z'\bar z$.
\begin{eqnarray*}
\dot{V} &= &\bar z'(A_V(\sigma(t))'+A_V(\sigma(t)))\bar z\\
&\leq& -2\lambda \bar z'\bar z
\end{eqnarray*}

Therefore, $\Phi_V(t,\tau)$  converges to zero as fast as $e^{-\lambda (t-\tau)}$ does, i.e.,
\[\|\Phi_V(t,\tau)\|\leq c e^{-\lambda(t-\tau)},\;\;\; \forall t\geq \tau\geq 0\]
$\qed$

\section{Conclusion}
This paper studies the distributed observer problem when the neighbor graph is time-varying but always strongly connected. 
It has been shown that for any switching signal with a dwell time or an average dwell time, for $g$ large enough, each agent can estimate the state exponentially fast with a pre-assigned convergence rate.
Study the distributed observer problem when the neighbor graph is not always strongly connected would be future work.

\bibliographystyle{unsrt}
\bibliography{lili, literature}

\begin{thebibliography}{10}

\bibitem{reza-cdc2005}
Reza Olfati-Saber.
\newblock {Distributed Kalman filter with embedded consensus filters}.
\newblock In {\em Proceedings of the 44th IEEE Conference on Decision and
  Control, and the European Control Conference, CDC-ECC '05}, pages 8179--8184,
  2005.

\bibitem{reza-cdc2007}
Reza Olfati-Saber.
\newblock {Distributed Kalman filtering for sensor networks}.
\newblock In {\em Proceedings of the 46th IEEE Conference on Decision and
  Control}, pages 5492--5498, 2007.

\bibitem{reza-tac2012}
Reza Olfati-Saber and Parisa Jalalkamali.
\newblock {Coupled distributed estimation and control for mobile sensor
  networks}.
\newblock {\em IEEE Transactions on Automatic Control}, 57(10):2609--2614,
  2012.

\bibitem{martins-cdc2012}
Shinkyu Park and Nuno~C. Martins.
\newblock {Necessary and sufficient conditions for the stabilizability of a
  class of LTI distributed observers}.
\newblock In {\em Proceedings of the 51st IEEE Conference on Decision and
  Control}, pages 7431--7436, 2012.

\bibitem{martins-tac2017}
Shinkyu Park and Nuno~C Martins.
\newblock Design of distributed lti observers for state omniscience.
\newblock {\em IEEE Transactions on Automatic Control}, 62(2):561--576, 2017.

\bibitem{trentlemen-scl2018}
W.~{Han}, H.~L. {Trentelman}, Z.~{Wang}, and Y.~{Shen}.
\newblock Towards a minimal order distributed observer for linear systems.
\newblock {\em Systems \& Control Letters}, 114:59 -- 65, 2018.

\bibitem{trentlemen-tac2019}
W.~{Han}, H.~L. {Trentelman}, Z.~{Wang}, and Y.~{Shen}.
\newblock A simple approach to distributed observer design for linear systems.
\newblock {\em IEEE Transactions on Automatic Control}, 64(1):329--336, Jan
  2019.

\bibitem{Mitra-tac2018}
Aritra Mitra and Shreyas Sundaram.
\newblock {Distributed Observers for LTI Systems}.
\newblock {\em IEEE Transactions on Automatic Control}, 63(11):3689--3704,
  2018.

\bibitem{Mitra-acc2019}
Aritra Mitra, John~A. Richards, Saurabh Bagchi, and Shreyas Sundaram.
\newblock Finite-time distributed state estimation over time-varying graphs:
  Exploiting the age-of-information.
\newblock In {\em Proceedings of the 2019 American Control Conference}, pages
  4006--4011, 2018.

\bibitem{CDC17}
L.~Wang, D.~Fullmer, A.~S. Morse, and J.~Liu.
\newblock A hybrid observer for a distributed linear system with a changing
  neighbor graph.
\newblock In {\em Proceedings of the 56th IEEE Conference on Decision and
  Control}, pages 1024--1029, Melbourne, Australia, Dec 2017.

\bibitem{TAC18}
L.~Wang and A.~S. Morse.
\newblock A distributed observer for a time-invariant linear system.
\newblock {\em IEEE Transactions on Automatic Control}, 63(7):2123--2130, 2018.

\bibitem{kim-cdc2016}
Taekyoo Kim, Hyungbo Shim, and Dongil~Dan Cho.
\newblock {Distributed Luenberger Observer Design}.
\newblock In {\em Proceedings of the 55th IEEE Conference on Decision and
  Control}, pages 6928--6933, Las Vegas, USA, 2016.

\bibitem{Kim-tac2019}
T.~{Kim}, C.~{Lee}, and H.~{Shim}.
\newblock Completely decentralized design of distributed observer for linear
  systems.
\newblock {\em IEEE Transactions on Automatic Control}, pages 1--1, 2019.

\bibitem{ACC19}
L.~Wang, J.~Liu, and A.~S. Morse.
\newblock A distributed observer for a continuous-time linear system.
\newblock In {\em Proceedings of 2019 American Control Conference}, pages
  86--89, Philadelphia, PA, USA, July 2019.

\bibitem{CDC19}
L.~Wang, J.~Liu, A.~S. Morse, and B.~D.~O. Anderson.
\newblock A distributed observer for a discrete-time linear system.
\newblock In {\em Proceedings of the 58th IEEE Conference on Decision and
  Control}, Nice, France, Dec 2019.

\bibitem{Morse-tac1996}
A.~S. {Morse}.
\newblock Supervisory control of families of linear set-point controllers -
  part i. exact matching.
\newblock {\em IEEE Transactions on Automatic Control}, 41(10):1413--1431, Oct
  1996.

\end{thebibliography}

\end{document}